%% file: main.tex
\documentclass[lettersize,journal]{IEEEtran}
\usepackage{xcolor}
\usepackage{colortbl} 
\definecolor{response_text}{RGB}{0, 0, 0}
\newcommand{\revone}[1]{{\color{response_text}#1}}
\usepackage{amsmath,amsfonts}
\usepackage{algorithmic}
\usepackage{algorithm}
\usepackage{array}
\usepackage[caption=false,font=footnotesize]{subfig}
\usepackage{textcomp}
\usepackage{stfloats}
\usepackage{url}
\usepackage{verbatim}
\usepackage{graphicx}
\usepackage{cite}
\usepackage{booktabs} 
\usepackage{multirow}
\usepackage{adjustbox}
\begin{document}

\title{Anonymization, Not Elimination: Utility-Preserved Speech Anonymization}

\author{Yunchong Xiao*, Yuxiang Zhao*, Ziyang Ma, Shuai Wang, 
        Kai Yu, Jiachun Liao, and Xie Chen%
\thanks{This work was supported by the National Natural Science Foundation of China  (No. U23B2018), Shanghai Municipal Science and Technology Major Project under Grant 2021SHZDZX0102 and Yangtze River Delta Science and Technology Innovation Community Joint Research Project (2024CSJGG1100).}
\thanks{Yunchong Xiao, Yuxiang Zhao, Ziyang Ma, Kai Yu, and Xie Chen are with the X-LANCE Lab, School of Computer Science, MoE Key Lab of Artificial Intelligence, Shanghai Jiao Tong University, Shanghai 200240, China (e-mail: {yunchongxiao, yuxiangzhao, zym.22, kai.yu, chenxie95}@sjtu.edu.cn).}%
\thanks{Shuai Wang is with the School of Intelligence Science and Technology, Nanjing University, Suzhou, China.}%
\thanks{Jiachun Liao is with the research center of big data technology, Nanhu Lab, Jiaxing, China.}%
\thanks{Yunchong Xiao and Yuxiang Zhao contributed equally to this work.}
\thanks{Jiachun Liao and Xie Chen are corresponding authors.}%
}

\markboth{Journal of \LaTeX\ Class Files,~Vol.~14, No.~8, August~2021}%
{Shell \MakeLowercase{\textit{et al.}}: A Sample Article Using IEEEtran.cls for IEEE Journals}
\maketitle
\input{Texts/0_abstract}
\begin{IEEEkeywords}
Speech anonymization, voice anonymization, speaker embedding generator, content anonymization, flow matching.
\end{IEEEkeywords}

\input{Texts/1_introduction}
\input{Texts/2_relatedwork}
\input{Texts/3_method}
\input{Texts/4_experiments}

\input{Texts/5_evaluation_setup}
\input{Texts/6_results}
\input{Texts/7_conclusion}
\bibliographystyle{IEEEtran}
\bibliography{refs}
\end{document}

%% file: Texts/0_abstract.tex
\begin{abstract}
The growing reliance on large-scale speech data has made privacy protection a critical concern. However, existing anonymization approaches often degrade data utility, for example by disrupting acoustic continuity or reducing vocal diversity, which compromises the value of speech data for downstream tasks such as Automatic Speech Recognition (ASR), Text-to-Speech (TTS), and Speech Emotion Recognition (SER). Current evaluation practices are also limited, as they mainly rely on direct testing of anonymized speech with pretrained models, providing only a partial view of utility. To address these issues, we propose a novel two-stage framework that protects both linguistic content and acoustic identity while maintaining usability. For content privacy, we employ a generative speech editing model to seamlessly replace personally identifiable information (PII), and for voice privacy, we introduce F3-VA, a flow-matching-based anonymization framework with a three-stage design that produces diverse and distinct anonymized speakers. To enable a more comprehensive assessment, we evaluate privacy using both acoustic- and content-based speaker verification metrics, and assess utility by training ASR, TTS, and SER models from scratch. Experimental results show that our framework achieves stronger privacy protection with minimal utility degradation compared to baselines from the VoicePrivacy Challenge, while the proposed evaluation protocol provides a more realistic reflection of the utility of anonymized speech under privacy protection.
\end{abstract}

%% file: Texts/1_introduction.tex
\section{Introduction}
The advent of foundation models has ushered in a data-centric era, where advanced speech systems for tasks such as Automatic Speech Recognition (ASR)~\cite{ahlawat2025automatic}, Text-to-Speech (TTS)~\cite{xie2025towards}, and Speech Emotion Recognition (SER)~\cite{george2024review} are increasingly trained on vast, internet-scale datasets of complex origin rather than well-curated academic corpora. While this paradigm shift has yielded unprecedented performance gains, it has also raised growing concerns regarding speech data privacy. Speech is an exceptionally rich data modality, containing not only explicit linguistic content but also implicit biometric markers capable of uniquely identifying an individual~\cite{backstrom2023privacy}. As a result, regulatory frameworks such as the General Data Protection Regulation (GDPR)~\cite{nautsch2019gdpr} explicitly classify voice recordings as sensitive biometric data, mandating strict legal compliance, explicit user consent, and robust security measures from data processors. Preserving speech privacy in a principled manner while ensuring that anonymized data remains suitable for model training has therefore emerged as a critical research imperative~\cite{miao2024synvox2}.
Privacy risks in speech data arise along two fundamental dimensions: content privacy and voice privacy. Content privacy concerns ``what is said'', whereas voice privacy concerns ``who is speaking''. Content privacy is primarily threatened by the extraction of Personally Identifiable Information (PII), such as names, addresses, or medical details, from ASR-transcribed text. \revone{Even in the absence of explicit identifiers, implicit linguistic patterns may still enable content re-identification or speaker linkage through characteristic stylistic cues, a risk that is particularly salient when speech is used as a medium for storage.} In contrast, voice privacy concerns biometric vocal characteristics that can be accurately recognized by Automatic Speaker Verification (ASV)~\cite{o2023review} systems, making it a primary concern in interactive scenarios.

\begin{figure}
\centering
\includegraphics[width=\columnwidth]{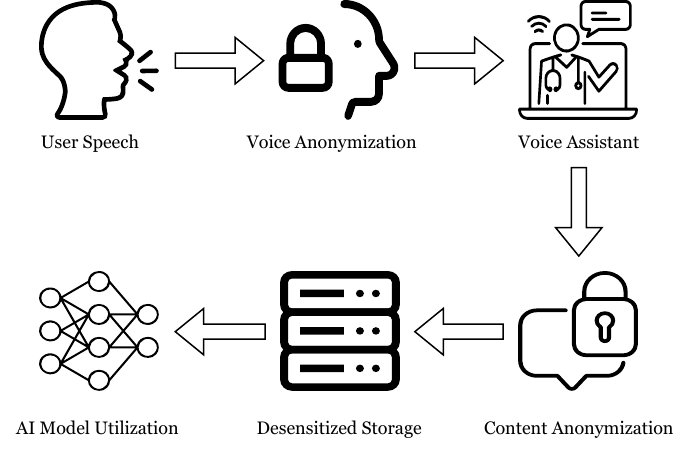}
\caption{Two-stage speech privacy protection framework: Voice anonymization preserves conversational content for interaction by obfuscating speaker identity, while content anonymization further modifies semantics for storage and model training.}
\label{fig_1}
\end{figure}

Fig.~\ref{fig_1} illustrates our two-stage framework for safeguarding speech privacy across the data lifecycle. In real-time applications such as online medical consultations, legal advisory services, and voice assistants, conversational fluency must be preserved while anonymizing the speaker’s voice to prevent identity leakage. When speech is stored for training or analysis, the focus shifts to anonymizing linguistic content to remove PII. This staged design addresses distinct privacy requirements at each phase without sacrificing the utility of anonymized data.

For voice anonymization, although initiatives such as the VoicePrivacy Challenge (VPC) series~\cite{tomashenko2022voiceprivacy,tomashenko2024voiceprivacy} have significantly advanced the field, most existing approaches rely on inductively biased feature disentanglement, in which speech is decomposed into speaker-related and speaker-unrelated representations for reconstruction~\cite{pierre22_interspeech}. Anonymization is typically achieved either by selecting substitute speakers from an external reference pool, which risks leaking the privacy of pool members, or by employing generative models such as Generative Adversarial Networks (GANs) and Variational Autoencoders (VAEs)~\cite{huang2024diffvc+,10022601}. \revone{However, limited diversity and controllability in such models often necessitate post-processing similarity checks to ensure sufficient separation from the original speaker.}

\revone{When it comes to content privacy, protection remains across multiple paradigms. Existing approaches typically follow cascaded designs in which ASR is first applied to obtain text and identify sensitive information, after which content protection is achieved through a range of mechanisms, including direct redaction of aligned speech segments~\cite{10363030}, semantic rewriting via large language models(LLMs) followed by speech resynthesis~\cite{dhingra2024enhancing}, or cryptography-based~\cite{7953391} schemes for secure storage and access control. While these cascaded pipelines are effective at mitigating explicit content leakage, they often introduce acoustic or semantic inconsistencies, or impose substantial constraints on data usability, thereby limiting their suitability for downstream model training.}

This paper fills a fundamental gap in the literature through the following contributions:
\begin{itemize}
    \item \textbf{Comprehensive Framework and In-depth Evaluation:} We propose a novel two-stage framework for protecting both voice and content privacy, paired with a more realistic evaluation protocol that assesses true data utility by training downstream models from scratch.
    \item \textbf{Novel Flow-Matching-based Voice Anonymization:} We introduce F3-VA, \revone{a speaker embedding-based voice anonymization model that generates diverse anonymized speaker embeddings while explicitly controlling the deviation from the original speaker embedding.}
    \item \textbf{A Generative Pipeline for Content Anonymization:} We propose Speech Editing Content Anonymizer (SECA), a complete pipeline that enhances traditional approaches by seamlessly replacing PII via generative speech editing while preserving acoustic integrity and fluency.
\end{itemize}

The remainder of this paper is organized as follows. Section II reviews related work. Section III details the proposed methodology. Section IV describes the experimental setup, followed by results and analysis in Section V. Section VI concludes the paper and discusses future work.

%% file: Texts/2_relatedwork.tex
\section{Related Work}

\subsection{Voice Anonymization}
The dominant paradigm in modern voice anonymization is founded on feature disentanglement~\cite{pierre22_interspeech}, wherein a speech signal is decomposed into distinct representations for linguistic content, prosody, and speaker identity. In practice, semantic content is often captured by features from self-supervised learning (SSL) models~\cite{baevski2020wav2vec,hsu2021hubert,chen2022wavlm}, prosody is represented by robust pitch extractors~\cite{kim2018crepe,wei23b_interspeech}, and speaker characteristics are captured by embeddings from pretrained ASV systems~\cite{8461375,desplanques20_interspeech,wang23ha_interspeech}. Anonymization is then achieved by manipulating or replacing the speaker representation. The central challenge within this paradigm lies in generating speaker identities that simultaneously preserve privacy and retain sufficient diversity, a challenge that has driven the evolution toward increasingly powerful generative models.


Early methods replace speaker embeddings with those from a reference pool~\cite{fang2019speaker,srivastava2020design,9829284}, which introduces privacy leakage from the reference speakers. Subsequent approaches attempt to mitigate this issue through embedding averaging~\cite{miao2023speaker,11026837,9247219} or by removing explicit speaker representations~\cite{10945405}, but often suffer from reduced speaker diversity and unrealistic identity modeling.

To remove reliance on external references altogether, later studies shifted toward directly generating speaker identities. VAEs have been used to model speaker embedding distributions~\cite{huang2024diffvc+}, while other approaches employed GAN to produce more diverse speaker embeddings~\cite{jafari2023speaker}. Nevertheless, these approaches typically generate random identities and require additional speaker selection or verification mechanisms~\cite{srivastava2020design} to ensure effective anonymization, increasing overall system complexity. \revone{More recently, flow matching~\cite{lipman2023flow} has emerged as a generative modeling paradigm with stable training dynamics and efficient sampling, providing a principled foundation for controllable speaker generation.}

\subsection{Content Anonymization}
Protecting the linguistic content of speech is equally critical, particularly in the era of foundation models where leakage of PII from training data poses significant risks for responsible AI deployment.

Traditional content anonymization methods follow cascaded pipelines based on ASR, Named Entity Recognition (NER), and localized redaction. Early studies investigated various infilling strategies, such as silence insertion, noise replacement, or speech reversal, to evaluate their impact on speech intelligibility and downstream utility~\cite{10363030}.

\revone{Building upon this paradigm, recent work has explored generative approaches that leverage LLMs to rewrite or paraphrase sensitive content at the text level, followed by speech resynthesis using TTS systems~\cite{dhingra2024enhancing,aggazzotti2025content}. While such pipelines produce semantically coherent and privacy-preserving utterances, they fundamentally regenerate the speech signal, discarding original acoustic characteristics, speaking style, and fine-grained prosodic patterns. Consequently, the resulting data become fully synthetic and may deviate substantially from the original distribution, limiting their suitability for downstream model training.}

\revone{An alternative line of research focuses on protecting speech content through cryptographic mechanisms rather than modifying the speech signal itself. In these approaches, speech waveforms or intermediate representations are encrypted to prevent unauthorized content extraction, while allowing controlled access under predefined conditions~\cite{tanaka2025voice,10.1561/116.20240020}. Client--server split architectures further support operations such as searchable phonetic queries over encrypted speech representations~\cite{7953391}. Although these methods provide strong content privacy guarantees without altering the original speech, they are primarily designed for secure storage or restricted access, rather than for generating openly usable anonymized datasets for model training.}

\revone{Recent advances in generative speech editing offer a complementary direction for content anonymization that aims to balance privacy protection and data utility. Traditional TTS-based replacement methods~\cite{jin2017voco} often introduce boundary artifacts and prosody inconsistencies at editing points. VoiceCraft~\cite{peng2024voicecraft} formulates speech editing as an autoregressive generation problem, which may lead to boundary discontinuities due to token rearrangement across edited segments. In contrast, F5-TTS adopts a non-autoregressive mask-prediction framework that supports multi-segment editing, but can exhibit prosody mismatch when the edited content differs substantially in duration from the original speech. While these approaches can effectively replace localized content, they generally preserve global voice characteristics, and prosodic inconsistencies become particularly pronounced in scenarios requiring fine-grained prosodic fidelity, such as pathological speech analysis~\cite{tayebi2024addressing,9648033}. Motivated by these limitations, we propose SECA, a generative content anonymization pipeline designed to seamlessly modify PII-bearing segments while better preserving speech naturalness and prosodic consistency.}
\subsection{Evaluation Paradigms for Speech Privacy Systems}

The evaluation of speech anonymization systems must jointly consider privacy protection and data utility. The VPC has established a widely adopted evaluation protocol, in which privacy is assessed via the performance degradation of speaker verification systems, while utility is measured by downstream task performance evaluated directly on anonymized speech signals.

Within the VPC framework, multiple attacker models are defined to reflect different levels of adversarial knowledge and capability:
\revone{\begin{itemize}
    \item \textbf{Ignorant Attacker:} the attacker is unaware of the anonymization process and performs speaker verification using enrollment data from the original speech domain.
    \item \textbf{Lazy-informed Attacker:} the attacker has access to the anonymization mechanism and applies the same transformation to enrollment data without further adaptation.
    \item \textbf{Semi-informed Attacker:} the attacker additionally exploits anonymized data to adapt or retrain attack models, representing a stronger and more realistic adversary.
\end{itemize}
These attacker models provide progressively stronger assessments of privacy robustness and are commonly used to benchmark voice anonymization systems under realistic threat assumptions.
}

\revone{Recent studies have shown that content re-identification can be achieved using linguistic content alone through stylistic and semantic cues, even when speaker-specific acoustic information is removed~\cite{gaznepoglu2025you}. Accordingly, content-based re-identification can be viewed as a form of speaker verification operating on content-related representations.}

Together, these considerations motivate a comprehensive evaluation paradigm for speech privacy systems that accounts for both acoustic- and content-based privacy risks, while also extending utility assessment beyond inference performance on anonymized speech. In particular, evaluating models trained from scratch on anonymized datasets enables a more faithful assessment of anonymized data as a training resource, capturing its impact on learnability, generalization, and long-term utility in realistic deployment scenarios.

%% file: Texts/3_method.tex
\section{Proposed methodology}
We propose a unified privacy protection framework for the speaker's timbre and content. 
This section will elaborate on three aspects: the overall voice anonymization framework, the core speaker anonymizer module, and the content anonymization module.
\input{Texts/3.1_speaker_anonymization}
\input{Texts/3.2_speaker_anonymizer}
\input{Texts/3.3_content_anonymization}

%% file: Texts/3.1_speaker_anonymization.tex

\subsection{Embedding-based Speech Reconstruction Backbone}

\begin{figure*}[htbp]
    \centering
    \subfloat[Training Architecture\label{fig:train}]{
        \includegraphics[height=9cm,keepaspectratio]{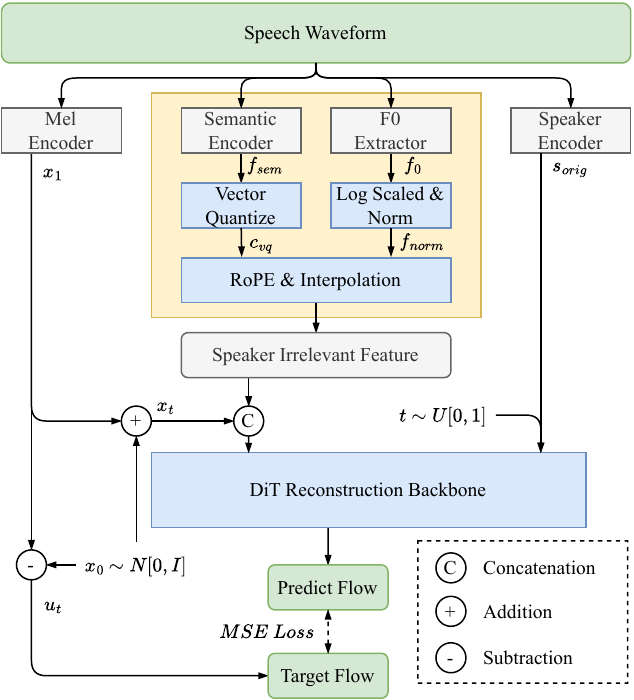}
    }
    \hfill
    \subfloat[Inference Architecture\label{fig:infer}]{
        \includegraphics[height=9cm,keepaspectratio]{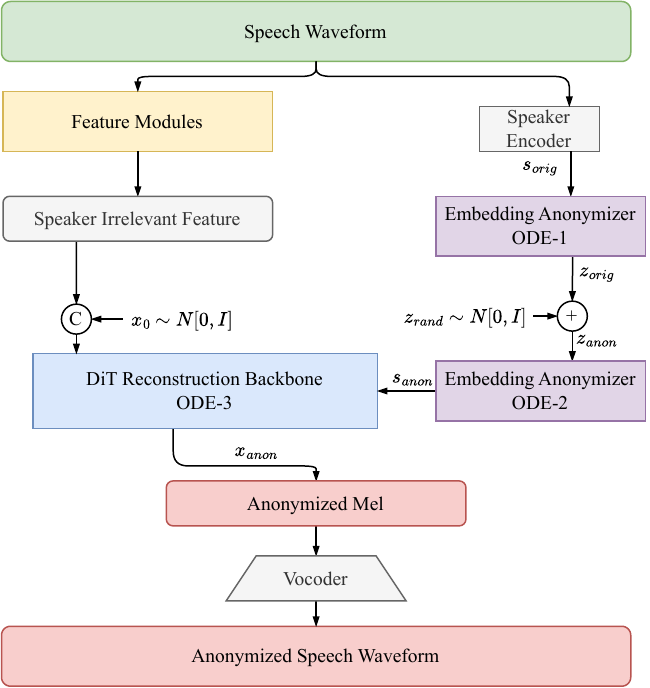}
    }
    \caption{\revone{
        Training and inference architecture.
        (a) Training phase: The backbone network predicts the flow based on the noisy mel, time step, and speaker irrelevant feature.
        (b) Inference phase: The reconstruction is performed by integrating the flow starting from Gaussian noise, conditioned on the anonymized speaker embedding generated by speaker anonymizer.
    }}
    \label{fig:training_inference}
\end{figure*}
The backbone network adopts an explicitly factorized reconstruction framework, in which speech is decomposed into distinct and interpretable representations—semantic content, prosodic features, and speaker identity—and then synthesized conditionally. This design is implemented using a conditional flow-matching model built upon a Diffusion Transformer (DiT)\cite{peebles2023scalable}, which has demonstrated strong performance in terms of training stability and generative diversity.

For linguistic content, we extract semantic features $f_{sem}$ using a pre-trained HuBERT-large model \footnote{\url{https://huggingface.co/facebook/hubert-large-ll60k}}. To remove residual speaker identity, we discretize these features through a vector quantization (VQ) layer \cite{gray1984vector} to obtain the quantized content representation $c_{vq}$:
\begin{equation}
\revone{c_{vq} = \text{VQ}(f_{sem})}.
\end{equation}
This quantization step serves as an information bottleneck to suppress residual speaker identity.

Prosodic features are derived from the fundamental frequency (F0) contour extracted using the RMVPE model\footnote{\url{https://huggingface.co/lj1995/VoiceConversionWebUI/tree/main}}. To obtain a speaker-independent representation, we compute the normalized pitch feature $p_{norm}$ by converting F0 to the semitone scale and subtracting the utterance-level median:
\begin{equation}
\revone{p_{norm} = s - \text{median}(s), \quad \text{where } s = 12 \cdot \log_2\left(\frac{f_0}{f_{\text{ref}}}\right),}
\end{equation}
and $f_{\text{ref}} = 440\,\text{Hz}$ is the standard reference pitch. Collectively, $c_{vq}$ and $p_{norm}$ constitute the speaker-irrelevant features used exclusively as conditioning inputs to the reconstruction backbone.

The speaker's identity is represented by a 192-dimensional speaker embedding $s_{orig}$ extracted from the pre-trained CAM++ model \footnote{\url{https://www.modelscope.cn/models/iic/speech_campplus_sv_zh_en_16k-common_advanced}}.

During training, the DiT-based backbone learns a probabilistic vector field (flow) $u_\theta$ to transform Gaussian noise $x_0 \sim \mathcal{N}(0, I)$ into the target mel-spectrogram $x_1$. We apply upsampling and Rotary Position Embeddings (RoPE)~\cite{su2024roformer} to align the speaker-irrelevant features ($c_{vq}$ and $p_{norm}$) with the mel-spectrogram.
As illustrated in Fig.~\ref{fig:training_inference}(a), at each time step $t \in [0, 1]$, we construct the noisy input $x_t = (1-t)x_0 + t x_1$.
This noisy input $x_t$ is concatenated along the channel dimension with the aligned speaker-irrelevant features to form the local input to the DiT. In contrast, the global speaker embedding $s_{orig}$ and time step $t$ are injected into every DiT block via the AdaLN-Zero mechanism, serving as global conditions to modulate the generation process.
The model predicts the flow, which is supervised by the target flow $u_t = x_1 - x_0$ via MSE loss.

%% file: Texts/3.2_speaker_anonymizer.tex

\subsection{Flow-Matching Speaker Embedding Anonymizer}

\begin{figure}[htbp]
    \centering
    \includegraphics[width=\columnwidth]{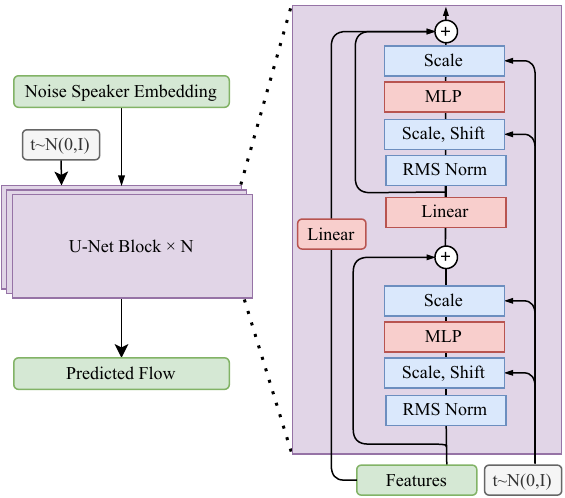}
    \caption{\revone{Training architecture of the flow-matching anonymizer. It employs a U-shaped network where standard attention layers are replaced by MLP blocks to process fixed-dimensional embeddings. The model takes a noisy embedding and time step as input to predict the flow.}}
    \label{fig:speaker_anonymizer}
\end{figure}
The speaker embedding anonymizer is a core component of our framework designed to generate anonymized speaker embeddings via a flow-matching generative model. 
As illustrated in Fig.~\ref{fig:speaker_anonymizer}, the anonymizer operates in the same flow-matching formulation as the backbone.
At each time step $t \in [0,1]$, the network takes a noisy speaker embedding $\phi_t(x)$ together with the time variable $t$ as input, and predicts the corresponding flow.
This flow specifies the instantaneous update direction of the embedding along the probability path between the Gaussian prior and the speaker embedding distribution.

To learn a reversible mapping between a simple Gaussian distribution and the speaker embedding distribution, we adopt the flow matching framework.
Let $p_1$ denote the distribution of real speaker embeddings and $p_0=\mathcal{N}(0,I)$ denote the standard Gaussian distribution.
The network $v_\theta$ defines the anonymizer ODE:
\begin{equation}
\frac{d}{dt}\,\phi_t(x) = v_\theta(\phi_t(x), t), \quad \phi_0(x) = x,
\label{eq:ode_def}
\end{equation}
\revone{where $\phi_t(x)$ represents the probability path under the Gaussian--speaker embedding distribution flow at time $t$, and $t \in [0,1]$ is the continuous time variable.}
The model is trained by minimizing the conditional flow matching loss which regresses the flow to the target path between the Gaussian distribution and the speaker embedding distribution.

\textbf{1) Speaker Encoding (ODE-1).}
Given a speaker embedding $s_{orig} \sim p_1$ extracted by CAM++, we encode it into the Gaussian distribution by solving the first ODE backward from $t=1$ to $t=0$:
\begin{equation}
z_{orig} = \phi_{0\leftarrow 1}(s_{orig})
\end{equation}
yielding $z_{orig} \sim \mathcal{N}(0,I)$.

\textbf{2) Speaker Obscuration.}
To anonymize the speaker, we sample a random noise $z_{\text{rand}} \sim \mathcal{N}(0,I)$ independently and combine it with the encoded variable $z_{\text{orig}}$, which is obtained deterministically by mapping $s_{\text{orig}}$ through the inverse flow:
\begin{equation}
z_{anon} = \frac{(1-w) \cdot z_{rand} + w \cdot z_{orig}}{\sqrt{(1-w)^2 + w^2}},
\label{eq:identity_obscuration}
\end{equation}
\revone{where $z_{anon}$ serves as the initial Gaussian noise for anonymized speaker embedding. Here, $w$ is a speaker weight controlling the strength of original speaker identity, $z_{orig}$ follows the standard Gaussian distribution through the learned flow, while $z_{rand}$ is independently sampled from $\mathcal{N}(0,I)$.
The normalization term $\sqrt{(1-w)^2 + w^2}$ ensures that the linear interpolation preserves unit variance.}

\textbf{3) Speaker Generation (ODE-2).}
We integrate the second ODE forward from $t=0$ to $t=1$ to map $z_{anon}$ back to the speaker embedding distribution:
\begin{equation}
s_{anon} = \phi_{1\leftarrow 0}(z_{anon})
\end{equation}

\textbf{4) Acoustic Reconstruction (ODE-3).}
The backbone ODE operates on the mel-spectrogram and defines a separate flow conditioned on the anonymized speaker embedding and speaker-irrelevant features.
Starting from Gaussian noise $x_0$, we generate the anonymized mel-spectrogram $x_{anon}$ by integrating the backbone flow:
\begin{equation}
\revone{x_{anon} = x_0 + \int_{0}^{1} u_\theta(x_t, t, s_{anon}, c_{vq}, p_{norm}) \, dt.}
\end{equation}
The resulting $x_{anon}$ is then synthesized into an anonymized speech waveform by the vocoder.

%% file: Texts/3.3_content_anonymization.tex
\subsection{Content Anonymization via Generative Editing}
\begin{figure}[htbp]
\centering
\includegraphics[width=\columnwidth]{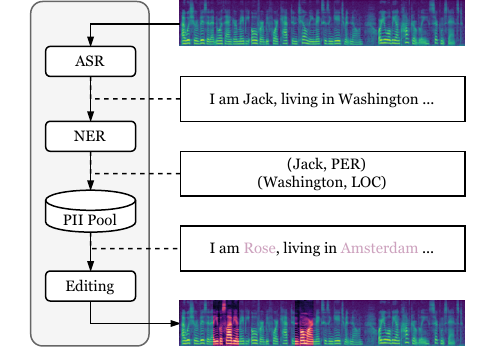}
\caption{A four-stage pipeline for content anonymization: 1) ASR, 2) NER, 3) replacement matching from a PII pool, and 4) generative speech editing.}
\label{fig:content_anonymization}
\end{figure}

To address the acoustic and semantic discontinuities caused by traditional redaction, our content anonymization module replaces sensitive information using a generative speech editing pipeline, as illustrated in Fig.~\ref{fig:content_anonymization}. 
The process begins by detecting PII in the transcription using the Flair NER model ner-english-large\footnote{\url{https://huggingface.co/flair/ner-english-large}}, which identifies four entity types: PER (person names), LOC (location names), ORG (organization names), and MISC (other names). Once PII is detected, a Connectionist Temporal Classification-based forced alignment tool\footnote{\url{https://docs.pytorch.org/audio/main/tutorials/ctc_forced_alignment_api_tutorial.html}} is used to provide frame-level alignment between the transcription and the speech signal, enabling accurate localization of PII segments in the speech.

In contrast to traditional methods that replace PII with silence or noise, our approach uses F5-TTS~\footnote{\url{https://huggingface.co/SWivid/F5-TTS}} for speech editing. 
To preserve utterance-level semantics and rhythm, we select replacement entities from the WikiAnn dataset~\footnote{\url{https://huggingface.co/datasets/unimelb-nlp/wikiann}} that match the entity type and length of the original PII, which helps preserve timbre and prosody consistency with the surrounding speech. Finally, the F5-TTS model updates the targeted speech segment based on the original text, the anonymized text, the original speech, and the time boundaries. 

ch

%% file: Texts/4_experiments.tex
\section{Experimental Setup}
This section details the experimental configuration used to evaluate our proposed framework, including the datasets, baseline models, and the evaluation protocol for both privacy and utility.
\input{Texts/4.1_datasets}
\input{Texts/4.2_baseline}

%% file: Texts/4.1_datasets.tex
\subsection{Datasets}

We conduct experiments on three widely-used open-source corpora: LibriSpeech \cite{7178964}, LibriTTS \cite{zen19_interspeech}, and IEMOCAP \cite{busso2008iemocap}, corresponding to ASR, TTS, and SER tasks, respectively. These benchmarks are selected to ensure the reproducibility and comparability of our results.

For ASR, we use the full LibriSpeech-train-960 to train downstream recognition models. For TTS, we use the full LibriTTS-train-554 for speech synthesis model training. For SER, we use the complete IEMOCAP dataset with a 5-fold cross-validation protocol. In addition, the voice anonymization model is trained on LibriSpeech, following the standard setup adopted in the VPC.

Beyond serving as training and test sets for downstream utility evaluation, these corpora are also used to construct dedicated trial sets for privacy assessment under the standard ASV trial protocol. Separate trial lists are generated for acoustic-based ASV (A-ASV) and content-based ASV (C-ASV) evaluations. A detailed summary of dataset usage, speaker statistics, and the number of trials for each subset is provided in Table~\ref{tab:dataset_summary}.

\input{Tables/1-dataset_usage}

\subsection{Implementation Details}
\subsubsection{Speech Reconstruction Backbone}
The core diffusion model adopts a DiT architecture with a hidden dimension of 768, a depth of 14 layers, and 12 attention heads. To enhance local information modeling, we replace the standard Feed-Forward Network blocks in DiT with ConvNeXtV2 blocks \cite{woo2023convnext}, where the hidden dimension is set to twice the DiT hidden dimension, and we introduce a long skip connection to improve training stability. The VQ layer has a dimensionality of 1024 and a codebook size of 1024. For feature processing, both semantic and F0 features are also transformed using ConvNeXtV2 blocks. Semantic features are mapped to a 512-dimensional representation, while semitone features are projected to 256 dimensions. The speech reconstruction backbone was trained on the LibriSpeech-600 for 500,000 steps across 8 NVIDIA RTX 3090 GPUs. The training objective is defined by a composite loss function that combines a vector quantization reconstruction loss $\mathcal{L}_{\text{commit}}$ and a flow matching Mean Squared Error loss $\mathcal{L}_{\text{flow}}$:

\begin{equation}
\mathcal{L}_{\text{total}} = \lambda \mathcal{L}_{\text{commit}} + \mathcal{L}_{\text{flow}}
\end{equation}

where the weight $\lambda$ is set to 1. Optimization is performed using the AdamW optimizer and a OneCycleLR scheduler with default parameters, except the pct\_start parameter set to 0.1.

\subsubsection{Speaker Embedding Anonymizer}

The speaker embedding anonymizer adopts a U-Net architecture tailored for fixed-length speaker embeddings. Given that speaker embeddings are non-sequential, the Multi-Head Self-Attention (MHSA) layers in the original DiT are replaced with lightweight Multi-Layer Perceptron (MLP) blocks. The input and output embeddings are 192-dimensional vectors, and the encoder-decoder structure performs dimensional scaling across six levels: \(192 \rightarrow 96 \rightarrow 48 \rightarrow 24 \rightarrow 48 \rightarrow 96 \rightarrow 192\), forming a symmetric U-shaped topology. The timestep embedding used for conditioning is also projected to 192 dimensions to match the latent space, ensuring consistent representation across modules. The embedding anonymizer was trained separately for 5000 epochs with a batch size of 128. The training was distributed across 8 NVIDIA RTX 3090 GPUs, utilizing the same AdamW optimizer and OneCycleLR scheduler as the backbone model. 

%% file: Tables/1-dataset_usage.tex
\begin{table*}[htbp]
\centering
\caption{Summary of Datasets Usage. A-ASV and C-ASV Denote Acoustic and Content-Based Automatic Speaker Verification, Respectively.}
\label{tab:dataset_summary}
\begin{adjustbox}{width=\textwidth}
\begin{tabular}{llccrr}  
\toprule
\textbf{Dataset}             & \textbf{Subset}                               & \textbf{\#Speakers}   & \textbf{Duration(hr)}    & \textbf{\#A-ASV Trials} & \textbf{\#C-ASV Trials} \\
\midrule
LibriSpeech-600              & train -clean-100 / -other-500                 & 1,731                 & 600               & N/A                        & N/A                               \\
LibriSpeech-960              & train -clean-100 / -clean-360 / -other-500    & 2,456                 & 960               & 9352                       & 8436                              \\
LibriSpeech-test-clean       & test-clean                                    & 40                    & 6.7               & 320                        & 256                              \\
LibriSpeech-test-other       & test-other                                    & 33                    & 8.6               & 264                        & 248                               \\
\midrule
LibriTTS-554                 & train -clean-100 / -clean-360 / other-500     & 2,311                 & 554               & 8880                       & 8432                              \\
LibriTTS-test-clean          & test-clean                                    & 39                    & 6.7               & 288                        & 240                               \\
LibriTTS-test-other          & test-other                                    & 33                    & 8.6                & 264                        & 248                               \\
\midrule
IEMOCAP                      & 5-fold cross-validation                       & 10                    & 7 hr                 & 200                        & 40                                \\
\bottomrule
\end{tabular}
\end{adjustbox}
\end{table*}

%% file: Texts/4.2_baseline.tex
\subsection{Baselines}
We compare our voice anonymization model F3-VA against two anonymization baselines from the VPC 2024. These two systems are also the final baselines retained in the VPC attacker track~\cite{tomashenko2024first}.

\begin{itemize}
    \item \textbf{NAC~\cite{10447871}:} This system, based on the Bark model, utilizes a Neural Audio Codec (NAC) language modeling approach. It generates anonymized speech by conditioning a GPT-like Transformer on semantic tokens from the original utterance and acoustic tokens from a pool of pseudo-speakers.
    \item \textbf{ASR-BN~\cite{champion2023anonymizing}:} This pipeline utilizes VQ bottleneck features from an ASR model that combines a pretrained wav2vec2 model with additional TDNN-F layers. A HiFi-GAN vocoder then synthesizes the anonymized waveform conditioned on these features and a target speaker identity one-hot vector.
\end{itemize}

\begin{table}[htbp]
\centering
\caption{\revone{Model Size and Runtime Efficiency of Anonymization Systems. Real-Time Factor Measured on a Single NVIDIA RTX 3090 GPU.}}
\label{tab:model_complexity}
\begin{tabular}{lcc}
\toprule
\textbf{Model} & \textbf{Model Parameters} & \textbf{Real-time Factor(RTF) $\downarrow$} \\
\midrule
NAC & 1221M & 1.62 \\
ASR-BN & 26M & 0.06 \\
F3-VA (Ours) & 180M & 0.23 \\
\bottomrule
\end{tabular}
\end{table}
\revone{Table~\ref{tab:model_complexity} summarizes the model size and real-time factor of the anonymization systems evaluated on a single NVIDIA RTX 3090 GP  with batch size=1. For F3-VA, the reported RTF is measured using the flow-matching inference process with 16 steps.}

%% file: Texts/5_evaluation_setup.tex
\section{Evaluation Setup}

\subsection{Privacy Evaluation}
Privacy is evaluated from two complementary perspectives: acoustic-based and content-based speaker verification. We use the Equal Error Rate (EER) as the privacy metric, where higher values indicate stronger anonymization. A-EER measures identity leakage through acoustic characteristics using an acoustic-based ASV system. C-EER measures identity leakage through linguistic content using linguistic re-identification system.

\textbf{Trial Construction:}
For each enrollment utterance, we construct two positive trials and two negative trials.
A positive trial pairs the enrollment utterance with another utterance from the same speaker, while a negative trial pairs it with an utterance from a different speaker.
Negative trials are balanced across genders by selecting one utterance from a male speaker and one from a female speaker.

For acoustic-based evaluation, candidate utterances are restricted to durations between 5 and 15 seconds. For content-based evaluation, only utterances whose transcripts contain PII are included, ensuring that content re-identification is evaluated on regions where content anonymization is actively applied.

\textbf{Speaker Verification Protocol:}
For acoustic-based evaluation, speaker embeddings are extracted using a pretrained ECAPA-TDNN model\footnote{\url{https://huggingface.co/speechbrain/spkrec-ecapa-voxceleb}}, and A-EER is computed based on the resulting verification scores. For content-based evaluation, we follow prior work on content re-identification~\cite{gaznepoglu2025you} and train a speaker identification model based on BERT for sequence classification\footnote{\url{https://huggingface.co/bert-base-uncased}} using textual transcriptions from the LibriSpeech and LibriTTS datasets, from which C-EER is computed.

Enrollment embeddings are computed at the speaker-level by averaging all utterance-level embeddings from the corresponding speaker, while trial embeddings are computed at the utterance level.
Cosine similarity between enrollment and trial embeddings is used as the verification score.

\subsection{Utility Evaluation}
To assess the utility of anonymized data, we evaluate its impact on three representative downstream tasks: ASR, TTS, and SER.
For each task, a state-of-the-art model is trained from scratch on anonymized training data and evaluated on the original, unmodified test set using the standard pipeline of its implementation.

\textbf{ASR Utility:}
We train a ZipFormer model using the official icefall recipe\footnote{\url{https://github.com/k2-fsa/icefall/tree/master/egs/librispeech/ASR/zipformer}} and report Word Error Rate (WER) from the official evaluation pipeline.

\textbf{TTS Utility:}
We train an F5-TTS model and evaluate it using the standard procedure of the F5-TTS repository.
We report WER for intelligibility, Speaker Embedding Cosine Similarity (SECS) for timbre preservation, and UTMOS~\cite{baba2024t05} for perceptual speech quality also from the official evaluation pipeline.

\textbf{SER Utility:}
We train an Emotion2Vec-based model\footnote{\url{https://github.com/ddlBoJack/emotion2vec/tree/main/iemocap_downstream}} and report Weighted Accuracy (WA), Unweighted Accuracy (UA), and the weighted average F1-score (F1).

In addition to downstream task performance, we report signal-level metrics to characterize intelligibility and perceptual quality of anonymized speech.
Intelligibility is assessed using the WER obtained with the Whisper-large-v3 model, and perceptual quality is evaluated using UTMOS

\revone{Regarding computational resources, all experiments were conducted on NVIDIA RTX 3090 GPUs. Training the ASR model required approximately 4 days using 4 GPUs, while training the TTS model required about 7 days using 8 GPUs. In contrast, the SER task was completed within a few hours on a single GPU.}

%% file: Texts/6_results.tex
\section{Results}
\begin{figure}[htbp]
\centering
\includegraphics[width=\columnwidth]{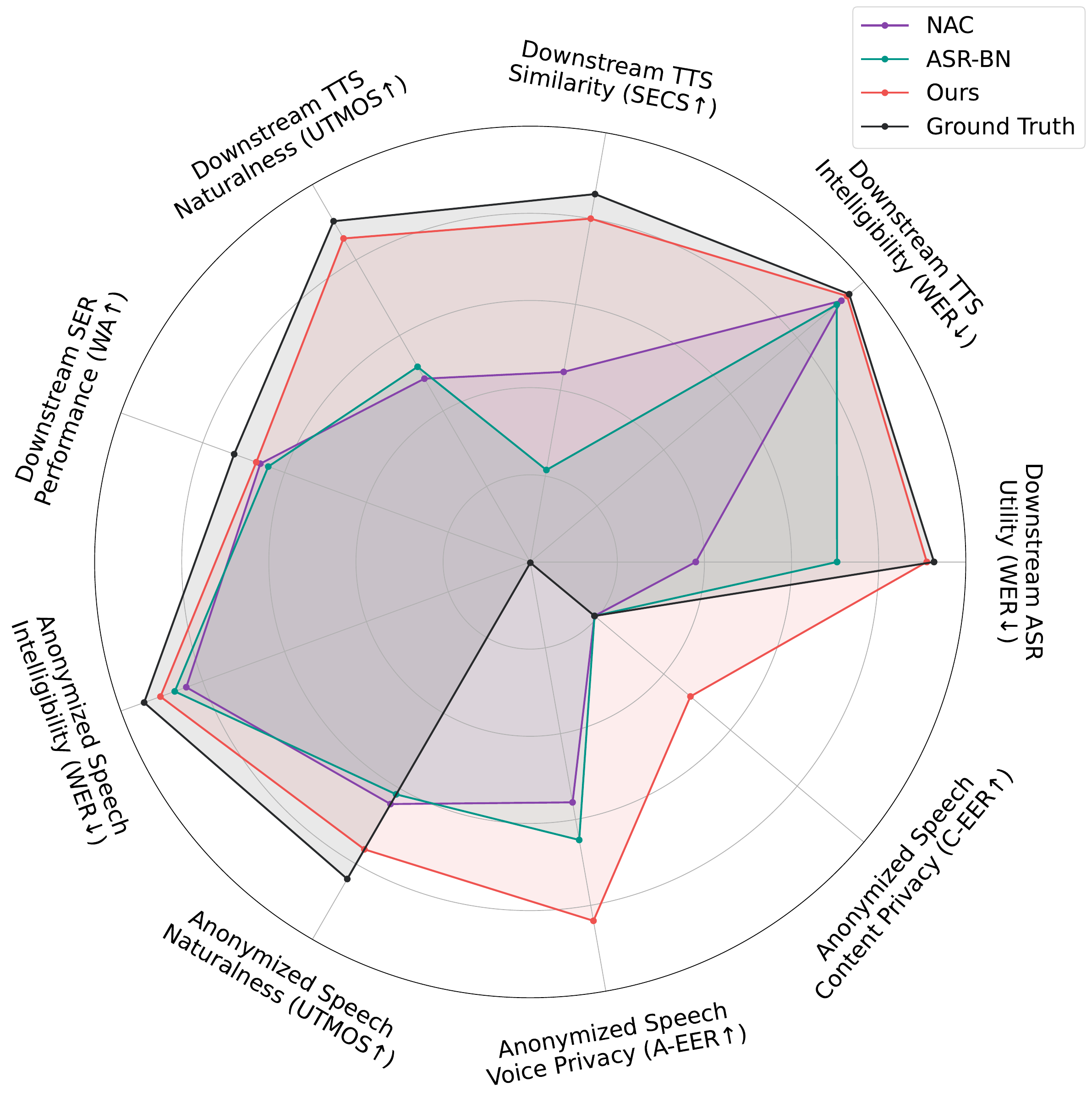}
\caption{\revone{Anonymization systems, assessing both direct speech quality metrics and the utility of anonymized data for training downstream tasks. Metrics are normalized to [0, 1] where 1 (outer edge) represents the best performance. The shaded area indicates the overall utility-privacy profile.}}
\label{fig:radar}
\end{figure}
To intuitively compare the models across multiple dimensions, we employ a Radar Chart. Since the evaluation metrics have different scales and interpretations, we normalize them onto a unified $[0, 1]$ scale where 1 consistently represents the best possible performance. For each metric, we define a reasonable performance range $[v_{\text{min}}, v_{\text{max}}]$. Specifically, these ranges are set to $[0\%, 50\%]$ for WER, $[1, 4.5]$ for UTMOS, $[0, 0.7]$ for SECS, $[0, 90\%]$ for WA, $[0\%, 75\%]$ for A-EER, and $[0\%, 50\%]$ for C-EER. The normalized score $v_{\text{norm}}$ is then computed as follows:

\begin{itemize}
    \item For "higher-is-better" metrics:
    \begin{equation}
        v_{\text{norm}} = \frac{v_{\text{raw}} - v_{\text{min}}}{v_{\text{max}} - v_{\text{min}}}
    \end{equation}

    \item For "lower-is-better" metrics:
    \begin{equation}
        v_{\text{norm}} = 1 - \frac{v_{\text{raw}} - v_{\text{min}}}{v_{\text{max}} - v_{\text{min}}}
    \end{equation}

\end{itemize}

The resulting normalized radar chart, depicted in Fig.~\ref{fig:radar}, allows for a direct and fair comparison where a larger enclosed area signifies superior overall performance. It provides a holistic comparison across the key evaluation dimensions, demonstrating that our proposed method (SECA + F3-VA) consistently outperforms the baselines with a performance profile that closely aligns with the Ground Truth. Conversely, the baselines exhibit significant trade-offs: ASR-BN shows a notable weakness in preserving speaker characteristics, whereas NAC significantly compromises linguistic content. The following subsections provide a detailed analysis of the metrics that underpin this visualization.

\subsection{Utility for Downstream Model Training}
\input{Tables/2-main_utility_privacy}

Table \ref{tab:utility_privacy} presents the core findings of our study, evaluating both privacy and utility through downstream task performance.
\revone{To substantiate these improvements and ensure they are not artifacts of random variance, we repeated the inference process using five distinct random seeds. All results reported below are averaged over these five runs.}

From a utility perspective, our voice anonymizer, F3-VA, preserves data utility, with downstream models trained on its output achieving performance close to the ground truth. For instance, the ASR model achieves a WER of 2.46\% on test-clean, only a minor degradation from 2.22\% on original data. Similarly, our content anonymizer, SECA, maintains high utility while targeting content privacy; the ASR WER is nearly identical to the ground truth (2.23\% vs. 2.22\%), and the TTS SECS remains high at 0.60. The NAC baseline severely degrades linguistic content (ASR WER of 21.00\%), while the ASR-BN baseline fails to preserve speaker diversity (TTS SECS of only 0.15).

From a privacy perspective, the results highlight the necessity of our two-stage approach. F3-VA achieves a high A-EER of 62.85\% on LibriSpeech. However, it does not address content-based re-identification, where the C-EER remains at the baseline level of 5.06\%. Conversely, SECA increases the C-EER to 17.80\% for ASR and 30.22\% for TTS, it means linguistic re-identification can partially rely on entity-level information for seen speakers. However, SECA offers no acoustic protection. Only the combined SECA + F3-VA system provides comprehensive protection across both acoustic and content dimensions.

\subsection{Analysis of Voice Anonymization}
Furthermore, we conduct direct performance evaluation on the anonymized speech to examine its surface-level quality degradation. As shown in Table \ref{tab:anonymized_data_metrics}, although our proposed model F3-VA consistently achieves the best results across both privacy and utility metrics, the performance gap between F3-VA and the two baselines is not always substantial. However, such direct evaluations alone are insufficient to fully reflect the suitability of anonymized data for downstream model training.
\input{Tables/3-direct_eval_utility_privacy}

For example, the NAC baseline exhibits only a modest increase in WER under direct evaluation, suggesting limited intelligibility loss. Yet, when this anonymized data is used to train an ASR model, the resulting performance degrades far more than the direct metrics would indicate.

In contrast, the ASR-BN baseline appears more balanced in most direct evaluation metrics compared to NAC. However, when used in the TTS task, its anonymized data yields a substantially lower SECS score of only 0.150, indicating poor consistency in timbre and speaker identity.

While direct metrics can assess speech quality and intelligibility directly, they fail to capture the deeper impact of anonymization on the learnability of the data. In contrast, our evaluation strategy of training the model directly on anonymized data can reveal hidden performance degradation and offers a more comprehensive and realistic assessment of utility.

\subsection{\revone{Analysis of Content Anonymization}}
\revone{To quantify the impact of cascading errors within the SECA pipeline, we conducted an ablation study on LibriSpeech-test-clean. We evaluated the system under two configurations: using ground-truth transcriptions to isolate editing quality (GT-Text), and adopting our Fully Cascaded SECA.}

\input{Tables/seca}

\revone{As detailed in Table~\ref{tab:seca_pipeline}, the localized editing approach maintains high acoustic fidelity, with UTMOS scores under both the GT-Text and Fully Cascaded settings remaining close to the Ground Truth. When the full SECA pipeline is applied, the word error rate increases to 4.80\%, and this degradation is primarily caused by upstream ASR inaccuracies that lead to pronunciation deviations in the edited segments. From a privacy perspective, the content-based EER increases from 24.75\% to 27.03\%, providing an important insight into the behavior of SECA. Although explicit PII is effectively removed through localized editing, the speaker’s linguistic style is largely preserved. This observation indicates that protecting against content-based re-identification requires mechanisms beyond simple entity redaction.}

\subsection{\revone{Analysis against Model-Aware Attackers}}
\input{Tables/attacker}
\revone{As shown in Table~\ref{tab:attacker}, for stochastic approaches, including the baselines and the proposed F3-VA with $w=0$, anonymized speaker representations are generated randomly or selected from a pool. The resulting mapping lacks deterministic structure, which prevents a Lazy-Informed attacker from exploiting model knowledge. Consequently, the performance gap between the Ignorant and Lazy-Informed attackers remains negligible, since the anonymized embedding does not deterministically depend on the original embedding.}

\revone{In contrast, when guidance is applied by setting the $w=-0.5$, a correlation is introduced between original and anonymized speaker embeddings. This configuration maximizes identity deviation against Ignorant attackers and achieves a high $EER_{ig}$ of 63.88\%. However, once the guidance rule is known, a Lazy-Informed attacker can exploit this determinism to realign embeddings, reducing the $EER_{la}$ to 35.28\%. These results indicate that although guided speaker weight yields higher peak privacy against ignorant adversaries, stochastic configurations offer stronger robustness against model-aware threats due to the irreversibility of the random transformation.}

\subsection{\revone{Analysis against Speaker Weight}}
\input{Tables/4-ablation_weight}
\revone{To further validate our conclusions, we study the effect of the speaker weight $w$. The analysis covers a wide range of values from $-1.0$ to $+1.0$, together with dynamic sampling strategies. For the Ignorant Attacker, the $EER_{ig}$ decreases monotonically from 67.92\% to 12.50\% as $w$ increases, demonstrating that $w$ directly controls the distance between the anonymized and original speakers.}

\revone{For the Lazy-Informed Attacker, deterministic guidance introduced by non-zero values of $w$ can be exploited. Under this attacker model, the $EER_{la}$ reaches its maximum of 52.08\% at $w=0$, where anonymization is purely stochastic, and decreases as the magnitude of guidance $|w|$ increases. The dynamic strategies exhibit consistent behavior: the range $[-1,1]$ behaves similarly to the stochastic baseline, while the range $[-1,0]$ provides a balanced trade-off comparable to moderate negative weights.}

\subsection{Analysis against Speaker Embedding Anonymizer}
\label{sec:ablation}

To further investigate the source of our model's superior performance, we conducted a focused ablation study on the TTS task, where speaker characteristics are most critical. We sought to answer the question: is our sophisticated speaker embedding anonymizer truly superior to the simpler approach of using just real voices from the training data? To this end, we compare two methods for generating anonymized speaker embeddings:

\begin{itemize}
    \item \textbf{Embedding Pool Selection}: This method simulates a common reference-pool-based technique. For each utterance, the original speaker embedding is replaced by a ground-truth embedding from a different, randomly selected speaker in the training set.
    \item \textbf{Speaker Anonymizer ($w=0$)}: This method uses our speaker embedding anonymizer with the speaker weight weight $w$ set to 0. This configuration ensures that the anonymizer generates a new embedding based solely on random noise, completely independent of the original speaker's identity.
\end{itemize}

The anonymized data from both methods was used to train a downstream TTS model. The results, presented in Table~\ref{tab:ablation_tts}, reveal a remarkable finding. The data anonymized by our F3-VA model yields a TTS model with a higher SECS (0.565 vs. 0.524) and a lower WER (2.12\% vs. 2.38\%) than the data anonymized using other ground-truth voices. This indicates that our speaker embedding anonymizer is not just replacing identities, but is capable of generating a \textit{more diverse} set of speaker embeddings than what is present in the original training data. This enhanced diversity, in turn, leads to a more robust and better-performing downstream TTS model, providing a strong explanation for the superior utility preservation demonstrated in our main results.

\begin{table}[ht!]
\centering
\caption{Ablation study on the speaker anonymizer, evaluated on the downstream TTS task.}
\label{tab:ablation_tts}
\resizebox{\columnwidth}{!}{%
  \begin{tabular}{@{}lccc@{}}
    \toprule
    \textbf{Method} & \textbf{TTS WER (\%) $\downarrow$} & \textbf{TTS SECS $\uparrow$} & \textbf{UTMOS $\uparrow$} \\
    \midrule
    Embedding Pool Selection & 2.38 & 0.524 & \textbf{4.00} \\
    \textbf{Speaker Anonymizer ($w=0$)} & \textbf{2.12} & \textbf{0.565} & 3.97 \\
    \bottomrule
  \end{tabular}%
}
\end{table}

\subsection{Visualization of Speaker Embedding Anonymization}
We visualized the speaker embedding space before and after anonymization to provide an intuitive, qualitative validation of our anonymization process. We randomly selected 10 speakers (5 male, 5 female) from the LibriSpeech-test set, with 100 utterances each. For each original speaker embedding, a corresponding anonymized embedding was generated using our speaker anonymizer with a speaker weight weight of $w=-0.5$. We then used t-SNE to project both sets of embeddings into a 2D space.

\begin{figure}[htbp]
    \centering
    \includegraphics[width=\columnwidth]{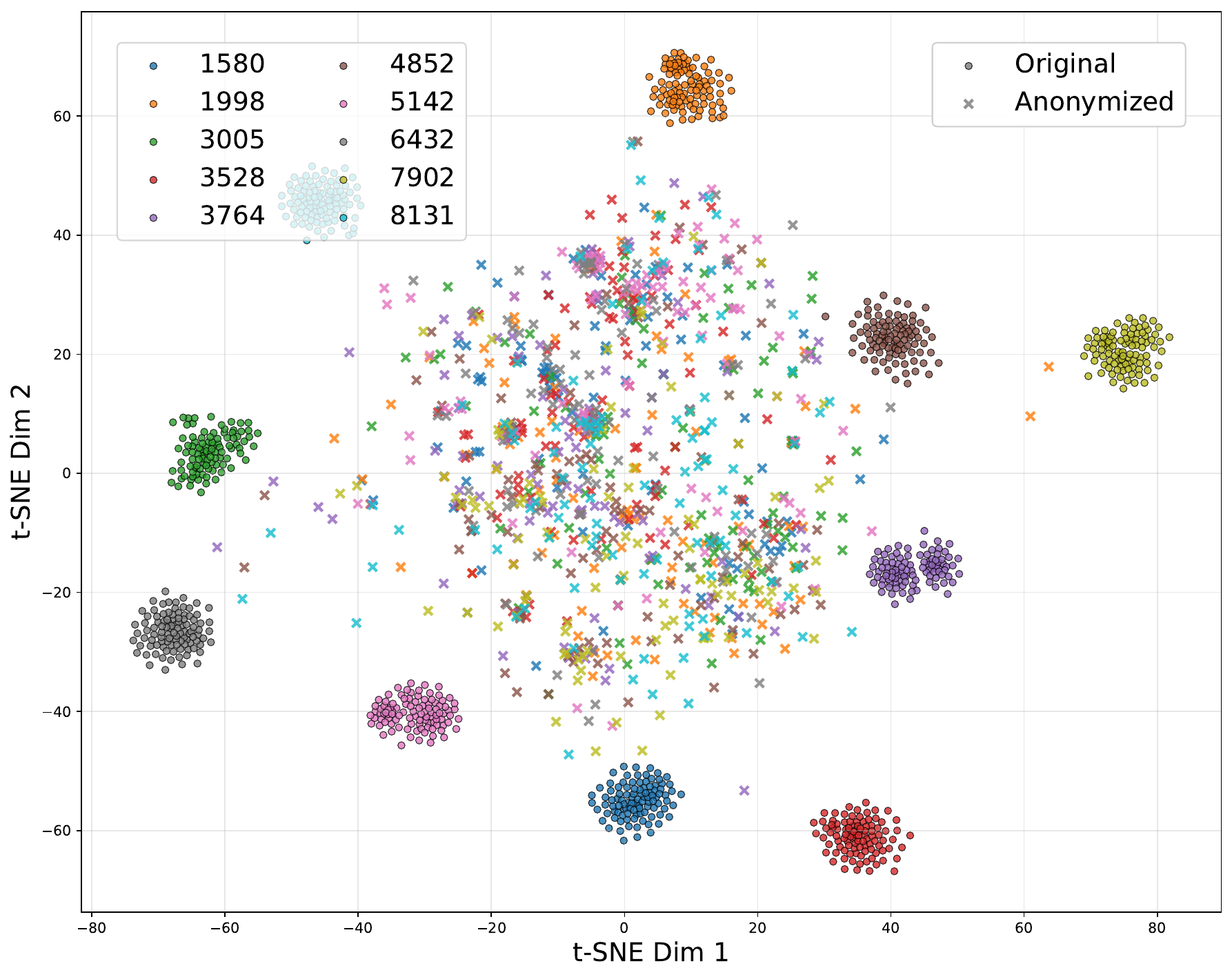}
    \caption{\revone{t-SNE visualization of speaker embeddings. Original embeddings (circles) form clusters, while anonymized embeddings (crosses) are dispersed.}}
    \label{fig:speaker_clustering}
\end{figure}

As depicted in Fig.~\ref{fig:speaker_clustering}, the original speaker embeddings, marked with circles, form tight, well-separated clusters, clearly distinguishing each of the 10 speakers. In contrast, the anonymized embeddings, marked with crosses, show no discernible clustering structure and are spread across the space in a quasi-uniform distribution. This visualization provides qualitative evidence that our method effectively erases speaker-discriminating information by converting original speaker embeddings into dispersed anonymized embeddings, breaking their original clustering structure and decorrelating them from their source identities.

%% file: Tables/2-main_utility_privacy.tex
\begin{table*}[htbp]
\centering
\caption{Utility and Privacy Results. WER, A-EER, C-EER, WA, UA, and F1 Are Reported in \%, While SECS and UTMOS Are Unitless. A-EER and C-EER Denote the Equal Error Rate for Acoustic-Based and Content-Based Speaker Verification, Respectively.}
\label{tab:utility_privacy}

\setlength{\tabcolsep}{2.8pt}
\renewcommand{\arraystretch}{1.05}

\begin{tabular}{@{}llllllllllllll@{}}
\toprule
\textbf{Tasks} 
& \multicolumn{3}{c}{\textbf{ASR}} 
& \multicolumn{5}{c}{\textbf{TTS}} 
& \multicolumn{5}{c}{\textbf{SER}} \\
\cmidrule(lr){1-1}               
\cmidrule(lr){2-4}
\cmidrule(lr){5-9}
\cmidrule(lr){10-14}

\textbf{Metrics} 
& \textbf{WER (clean/other)$\downarrow$} & \textbf{A-EER$\uparrow$} & \textbf{C-EER$\uparrow$}
& \textbf{WER$\downarrow$} & \textbf{SECS$\uparrow$} & \textbf{UTMOS$\uparrow$} 
& \textbf{A-EER$\uparrow$} & \textbf{C-EER$\uparrow$}
& \textbf{WA$\uparrow$} & \textbf{UA$\uparrow$} & \textbf{F1$\uparrow$} 
& \textbf{A-EER$\uparrow$} & \textbf{C-EER$\uparrow$} \\
\midrule

Ground Truth 
& 2.22/5.08 & 0.13 & 5.06 
& 2.20 & 0.60 & 4.16 & 0.14 & 14.21 
& 72.05 & 72.87 & 72.05 & 1.00 & 44.44 \\

\midrule
NAC 
& 21.00/41.00 & 42.00 & 5.06 
& \underline{3.37} & \underline{0.31} & 2.70 & 42.02 & 14.21 
& \underline{65.57} & \underline{66.51} & \underline{65.71} & \underline{44.00} & 44.44 \\

ASR-BN 
& \underline{5.08}/\underline{24.48} & \underline{48.76} & 5.06 
& 4.07 & 0.15 & \underline{2.81} & \underline{48.40} & 14.21 
& 63.51 & 64.94 & 63.51 & \underline{53.00} & 44.44 \\

F3-VA (Ours) 
& \textbf{2.46}/\textbf{5.98} & \textbf{62.85} & 5.06 
& \textbf{2.22} & \textbf{0.56} & \textbf{3.99} & \textbf{62.82} & 14.21 
& \textbf{67.13} & \textbf{67.94} & \textbf{67.21} & \textbf{57.00} & 44.44 \\

SECA (Ours) 
& 2.23/5.28 & 0.26 & \textbf{17.80} 
& 2.41 & 0.60 & 4.10 & 0.13 & \textbf{30.22} 
& 71.68 & 72.51 & 71.64 & 1.60 & \textbf{58.33} \\

\textbf{SECA + F3-VA} 
& 2.60/6.35 & 62.51 & \textbf{17.80} 
& 2.53 & 0.56 & 4.00 & 62.51 & \textbf{30.22} 
& 66.61 & 67.08 & 67.08 & 56.00 & \textbf{58.33} \\

\bottomrule
\end{tabular}
\end{table*}

%% file: Tables/3-direct_eval_utility_privacy.tex
\begin{table}[htbp]
\centering
\caption{Performance Comparison of Models Across Anonymized Data.}
\label{tab:anonymized_data_metrics}
\resizebox{\columnwidth}{!}{
\begin{tabular}{llccc}
\toprule
\textbf{Test Set} & \textbf{Model} & \textbf{EER\%$\uparrow$} & \textbf{WER\%$\downarrow$} & \textbf{UTMOS$\uparrow$} \\
\midrule


\multirow{4}{*}{LibriSpeech-test-clean}
& Ground Truth       & 0.00 & 2.06 & 4.13 \\  
& ASR-BN            & \underline{45.63} & \underline{4.64} & 3.24 \\
& NAC               & 41.88 & 6.00 & \underline{3.33} \\
& F3-VA           & \textbf{63.13} & \textbf{3.01} & \textbf{3.72} \\
\midrule

\multirow{4}{*}{LibriSpeech-test-other}
& Ground Truth       & 0.00 & 3.25 & 3.53 \\  
& ASR-BN            & \underline{45.46} & \underline{12.52} & 3.14 \\
& NAC               & 37.88 & 14.51 & \underline{3.18} \\
& F3-VA           & \textbf{55.30} & \textbf{7.54} & \textbf{3.68} \\
\midrule


\multirow{4}{*}{LibriTTS-test-clean}
& Ground Truth       & 0.00 & 1.67 & 4.16 \\  
& ASR-BN            & \underline{50.00} & \underline{4.80} & 3.25 \\
& NAC               & 43.06 & 6.59 & \underline{3.29} \\
& F3-VA           & \textbf{64.58} & \textbf{2.78} & \textbf{3.68} \\
\midrule

\multirow{4}{*}{LibriTTS-test-other}
& Ground Truth       & 0.00 & 2.91 & 3.54 \\  
& ASR-BN            & 46.97 & \underline{11.17} & 3.11 \\
& NAC               & \underline{43.18} & 13.13 & \underline{3.19} \\
& F3-VA           & \textbf{61.36} & \textbf{6.36} & \textbf{3.66} \\
\midrule

\multirow{4}{*}{IEMOCAP}
& Ground Truth       & 1.00 & 10.96 & 2.36 \\  
& ASR-BN            & 44.00 & \underline{26.88} & \underline{2.56} \\
& NAC               & \underline{53.00} & 28.51 & 2.51 \\
& F3-VA           & \textbf{57.00} & \textbf{17.54} & \textbf{2.99} \\
\bottomrule
\end{tabular}
}
\end{table}

%% file: Tables/seca.tex
\begin{table}[htbp]
    \centering
    \caption{\revone{Analysis of Cascading Errors in the SECA Pipeline.}}
    \label{tab:seca_pipeline}
    \revone{
        \begin{tabular}{lccc}
            \toprule 
            Pipeline & WER (\%) \textdownarrow & UTMOS \textuparrow & C-EER (\%) \textuparrow \\
            \midrule
            Ground Truth & 2.48 & 4.10 & 24.75 \\
            SECA(GT-Text) & 4.27 & 4.06 & 27.03 \\
            SECA (Fully Cascaded) & 4.80 & 4.05 & 26.80 \\
            \bottomrule
        \end{tabular}
    }
\end{table}

%% file: Tables/attacker.tex
\begin{table}[htbp]
    \centering
    \caption{\revone{Analysis Against Ignorant ($\mathrm{EER}_{\mathrm{ig}}$) and Lazy-Informed ($\mathrm{EER}_{\mathrm{la}}$) Attackers Under Different Speaker Weights.}}
    \label{tab:attacker}
    
    \revone{
        \begin{tabular}{lcccc} 
            \toprule
            \textbf{Model} & \textbf{$\mathrm{EER}_{\mathrm{ig}}$\%}$\uparrow$ & \textbf{$\mathrm{EER}_{\mathrm{la}}$\%}$\uparrow$ & \textbf{WER\%}$\downarrow$ & \textbf{UTMOS}$\uparrow$ \\
            \midrule
            ASR-BN              & 50.00 & \textbf{47.50} & 5.40 & 3.27 \\
            NAC                 & 45.00 & 45.00 & 6.44 & 3.21 \\
            F3-VA ($w=0.0$)     & 45.46 & 43.75 & 3.02 & \textbf{3.88} \\
            F3-VA ($w=-0.5$)    & \textbf{63.88} & 35.28 & \textbf{2.83} & 3.86 \\
            \bottomrule
        \end{tabular}
    }
\end{table}

%% file: Tables/4-ablation_weight.tex
\begin{table}[htbp]
    \centering
    \caption{\revone{Privacy and Utility Results Across Fixed and Dynamic Speaker Weight Strategies.}}
    \label{tab:updated_robustness}

    \revone{
        \begin{tabular}{ccccc}
            \toprule
            \textbf{Speaker Weight} &
            \textbf{$\mathrm{EER}_{\mathrm{ig}}$\%}$\uparrow$ &
            \textbf{$\mathrm{EER}_{\mathrm{la}}$\%}$\uparrow$ &
            \textbf{WER\%}$\downarrow$ &
            \textbf{UTMOS}$\uparrow$ \\
            \midrule
            $-1.00$ & \textbf{67.92} & 19.58 & 3.36 & 3.84 \\
            $-0.75$ & 64.58 & 28.75 & 3.34 & 3.85 \\
            $-0.50$ & 65.00 & 37.92 & 4.03 & 3.88 \\
            $-0.25$ & 52.92 & 40.00 & 3.73 & 3.85 \\
            \midrule
            $0.00$  & 45.00 & \textbf{52.08} & 3.15 & 3.87 \\
            \midrule
            $0.25$  & 35.00 & 45.00 & 3.64 & 3.85 \\
            $0.50$  & 27.50 & 29.58 & 3.91 & 3.82 \\
            $0.75$  & 20.00 & 19.17 & 3.65 & 3.83 \\
            $1.00$  & 12.50 & 13.75 & 3.48 & 3.84 \\
            \midrule
            $[-1, 1]$   & 45.83 & 43.75 & \textbf{2.92} & 3.87 \\
            $[-1, 0]$   & 61.25 & 36.25 & 3.84 & \textbf{3.89} \\
            \bottomrule
        \end{tabular}
    }
\end{table}

%% file: Texts/7_conclusion.tex
\section{Conclusion}
In this paper, we presented and validated a comprehensive two-stage framework for speech privacy protection that preserves data utility for downstream tasks while protect both voice privacy and elementary content privacy. By evaluating utility through models trained from scratch on anonymized data, we demonstrated a more realistic assessment of the value of anonymized speech. The proposed framework integrates SECA for content anonymization and F3-VA for voice anonymization, and experimental results show that it enables effective privacy protection while supporting high-performing ASR, TTS, and SER models.

\revone{Looking forward, speech anonymization will continue to face increasingly strong and informed adversaries as anonymization models become more capable. Future studies should therefore pay closer attention to privacy robustness under advanced threat models, especially for content anonymization, where content re-identification can emerge from linguistic style and semantic patterns. Addressing these challenges will likely require broader perspectives, potentially drawing on ideas from large language models, cryptography, and other principled privacy frameworks. At the same time, practical deployment calls for robustness under real-world conditions, such as background noise, conversational speech, multilingual speech, and diverse recording environments. Progress along these directions is essential for developing scalable and reliable privacy-preserving speech systems.}